\documentclass[a4paper,12pt,times,authoryears]{amsart}

\usepackage{geometry}
\usepackage{lipsum}
\usepackage{amsaddr}

 \makeatletter
\renewcommand{\email}[2][]{%
  \ifx\emails\@empty\relax\else{\g@addto@macro\emails{,\space}}\fi%
  \@ifnotempty{#1}{\g@addto@macro\emails{\textrm{(#1)}\space}}%
  \g@addto@macro\emails{#2}%
}
\makeatother

\usepackage{bm}
\usepackage{amssymb}
\usepackage{graphicx}
\usepackage{mathrsfs}
\usepackage{subfig}
\usepackage{float}
\usepackage{natbib}
\usepackage{booktabs}
\usepackage{epstopdf}

\usepackage[colorlinks,bookmarksopen,bookmarksnumbered,citecolor=red,urlcolor=red]{hyperref}

\numberwithin{equation}{section}

  \def\bD{{\mathbf D}} 
  
   \def\bI{{\mathbf I}}

  \def\bX{{\mathbf X}} \def\bY{{\mathbf Y}}

\def\bbeta{{\boldsymbol{\beta}}}

 \def\veps{\varepsilon}

\def\real{\mathop{{\rm I}\kern-.2em\hbox{\rm R}}\nolimits}

\def\1overn{\frac{1}{n}}

\def\bel{\begin{eqnarray}\label}  \def\eel{\end{eqnarray}}
\def\bes{\begin{eqnarray*}}  \def\ees{\end{eqnarray*}}

\theoremstyle{definition}

\theoremstyle{remark}

\newcommand{\ed}{\end{document}}

\begin{document}

\title[Big Data Analysis Using Shrinkage Strategies]{Big data analysis using shrinkage strategies}

\author{Bahad{\i}r Y\"{u}zba\c{s}{\i}}
\address[Bahad{\i}r Y\"{u}zba\c{s}{\i}]{Department of Econometrics, Inonu University, Turkey}
\email[Bahad{\i}r Y\"{u}zba\c{s}{\i}]{b.yzb@hotmail.com}

\author{Mohammad Arashi}
\address[Mohammad Arashi]{Department Statistics, Shahrood University of Technology, Shahrood, Iran}
\email[Mohammad Arashi]{m\_arashi\_stat@yahoo.com}

\author{S. Ejaz Ahmed}
\address[S. Ejaz Ahmed]{Department of Mathematics and Statistics, Brock University, Canada}
\email[S. Ejaz Ahmed]{sahmed5@brocku.ca}

\date{}

\begin{abstract}
In this paper, we apply shrinkage strategies to estimate regression coefficients efficiently for the high-dimensional multiple regression model, where the number of samples is smaller than the number of predictors. We assume in the sparse linear model some of the predictors have very weak influence on the response of interest. We propose to shrink estimators more than usual. Specifically, we use integrated estimation strategies in sub and full models and shrink the integrated estimators by incorporating a bounded measurable function of some weights. The exhibited double shrunken estimators improve the prediction performance of sub models significantly selected from existing Lasso-type variable selection methods. Monte Carlo simulation studies as well as real examples of eye data and Riboavin data confirm the superior performance of the estimators in the high-dimensional regression model.

\medskip

\noindent \text{Keywords:} Double shrinkage; High-dimension; Penalty; Prediction; Sparse regression model.

\end{abstract}

\maketitle

\section{Introduction}
\noindent
Nowadays many researchers have focused on the analysis of big data, because of existence trend in computer science and statistics. Comparing to the usual datasets, big data refer to high-dimensional, unusual and unstructured data. The analysis of big data needs methods other than traditional analytical frameworks.

Let $n$ denote the sample size or number of observations and $p$ the number of features or variables. Data scientists consider big data as when $n$ is too large, however in medical and genetic researches, engineering and financial studies, one mostly involves small $n$ large $p$ problem, known as high-dimensional data. As \citet{Pyne-et-al} pointed in big data analytics, some domains of big data such as finance or health do even produce infinite dimensional functional data, which are observed not as points but functions, such as growth curves, online auction bidding trends, etc. As \citet{Wang-et-al} pointed, big data are data on a massive scale in terms of volume, intensity, and complexity that exceed the capacity of standard analytic tools. \citet{Ahmed2014b} collected some research contributions in the field of big data analytics to highlight high-dimensional methods in big data challenges.

As \cite{AhmedYuzbasi2016} and \cite{AhmedYuzbasi2016b} pointed, the term ``big data" is not well defined, but its problems are real and statisticians need to play a more important role in this arena. The big data or data science is an emerging field. In 2013, American Statistical Association (ASA) proposed three reasons to show ASA has not been very involved in big data. President of Institute of Mathematical Statistics (IMS), Bin Yu, in 2014 called for statisticians to own data science by working on real problems such as those from genomics, neuroscience, astronomy, nanoscience, computational social science, personalized medicine/healthcare, finance, and government; relevant methodology/theory will follow naturally.

There is an increasing demand for efficient prediction strategies for analyzing high-dimensional data in big data streams. For example, data arising from gene expression arrays, social network modeling, clinical, genetics and phenotypic data. Due to the trade-off between  model complexity and model prediction, the statistical inference of model selection becomes an extremely important and challenging problem in high-dimensional data analysis. Over the past two decades, many penalized regularization approaches have been
 developed  to do variable selection and estimation simultaneously.
 Among them,  least absolute shrinkage and selection operator (LASSO) is one of the recent
 approaches, \citet{Tibshirani1996}. It is a useful technique due to its convexity and computation efficiency. The LASSO is based on squared error and a penalty proportional to regression parameters. \citet{Schelldorfer2011} provides a comprehensive summary of the consistency properties of the LASSO. \citet{Efron2004} introduced the least angle regression algorithm which is a very fast way to draw the entire regularization path for a LASSO estimate of the regression parameters. The penalized likelihood methods have been extensively studied in the literature, see for example, \citet{Tran2011}, \citet{Huang2008}, \citet{Kim2008}, \citet{Wang2007}, \citet{Yuan2006}, \citet{Leng2006}, and \citet{Tibshirani2005}. The penalized likelihood methods have a close connection to Bayesian procedures. Thus the LASSO estimate corresponds to a Bayes method that puts a Laplacian (double exponential) prior on the regression coefficients. Recent results (\citet{Armagan2013},   \citet{Bhattacharya2012}, and   \citet{Carvalho2010}) have demonstrated that better desirable results can be obtained by using priors with heavier tails than the double exponential prior, in particular, priors with polynomial tails. Our study has concentrated on the widely recognized penalty estimators LASSO and adaptive LASSO (ALASSO). Very recently, \cite{YuzbasiArashi2016} have proposed double shrinking concept to improve the prediction accuracy of LASSO. Here, we specifically implement the double shrunken estimator on ALASSO.

Following \cite{AhmedYuzbasi2016}, we consider the estimation problem of regression parameters when there are many potential predictors in the initial/working model and:
 \begin{enumerate}
   \item most of them may not have any influence (sparse signals) on the response of interest
   \item some of the predictors may have strong influence (strong signals) on the response of interest
   \item some of them may have weak-moderate influence (weak-moderate signals) on the response of interest
 \end{enumerate}

It is possible that there may be extraneous predictors in the model. Suppose if the main concern is treatment effect, or the effect of biomarkers, extraneous nuisance variables may be lab effects when several labs are involved, or the age and sex of patients. The analysis will be more precise if ``nuisance variables'' can be left out of the model.
This leads to the consideration of two models: the full model that includes all predictors and possible extraneous variables, and a candidate submodel that includes the predictors of main concern while leaving out extraneous variables. Further, it is important that we do not automatically remove all the predictors with weak signals from the model.  This may result in selecting a biased submodel. A logical way to deal with this framework is to use pretest model selection and estimation strategies that test whether the coefficients of the extraneous variables are zero and then estimate parameters in the model that include coefficients that are rejected by the test. Another strategy is to use Stein-type shrinkage estimators where the estimated regression coefficient vector is shrunk in the direction of the candidate subspace. This ``soft threshold'' modification of the pretest method has been shown to be efficient in various frameworks. \citet{Ahmed2012},  among others have investigated the properties of shrinkage and pretest methodologies for host models.

The model and some estimators are introduced in Section 2. In Section 3 we showcase our suggested estimation strategy. The results of a simulation study that includes comparison of suggested estimator with the penalty estimators are reported in Section 4. Application to real data sets is given in Section 5. Finally, we offer concluding remarks in Section 6.

\section{Estimation strategies}

In this communication, we consider a high-dimensional linear regression sparse model:

\begin{equation}
y_i=\sum_{j=1}^{p} x_{ij} \beta_j+\veps_i,\quad 1\le i\le n << p
\end{equation}
where $y_i$ observed
response variable with predictors $x_i$s, and $\beta_j$ are the regression parameters. Further,
$\veps_i$s are  independent and identically distributed random errors
with center $0$ and variance $\sigma^2$. Similar to most of LASSO penalty-type models, in our approach, we assume the true model is sparse in the sense that most of regression coefficients are zeros except for a few ones and all nonzero $\beta_j$'s are larger than noise level,  $c\sigma\sqrt{(2/n)\log(d)}$ with $c\ge 1/2$.
We refer to \citet{Zhao2006}, \citet{Huang2008}, and \citet{Bickel2009} for some insights.
In general, the LASSO penalty turns to select an
over-fitted model since it penalizes all coefficients equally (\citet{Leng2006}). In reviewed literature several modification and methodologies have been suggested to improve the prediction accuracy for LASSO strategy. For example, the SCAD (\citet{FanLi2001}), adaptive LASSO, (\citet{Zou2006}), MCP (\citet{Zhang2010}) and Stein-type LASSO (\cite{YuzbasiArashi2016})
 and several others. These methods select a submodel by shrinking some regression coefficients to zero and provide shrinkage estimators of the remaining coefficients. However, these methods may force the relatively more weak coefficients towards zeros as compared to LASSO, resulting in under-fitted models subject to a much larger selection bias in the presence of significant number of weak signals.

Following \cite{AhmedYuzbasi2016} and \cite{AhmedYuzbasi2016b}, in this paper, we consider the estimation and prediction problem for the sparse regression models when there are many potential predictors that have weak influence on the response of interest. The analysis will be relatively more precise if ``weak effect" variables can be weighted for the ultimate model prediction. This leads to the consideration of two models: the over-fitted model that includes predictors with strong signals and possibly some predictors with weak signals selected by LASSO.  On the other, we select an underfitted model that possibly includes the predictors with strong signals while leaving out predictors with weak effect by using ALASSO. One way to deal with this framework is to use Stein-type shrinkage estimators where the estimated regression coefficient vector is shrunk in the direction of the under-fitted model. This ``soft threshold'' modification of the pretest method has been shown to be efficient in various frameworks. \citet{Saleh2006} and \citet{Ahmed2012},  among others have examined the properties of Stein-type shrinkage estimation strategies for a host of models.

Consider the following regression model
\begin{equation}\label{eq:ch1:full:model}
\bY=\bX_n \bm\beta+ \bm\varepsilon,
\end{equation}
where $\bY=(y_1, y_2, \dots, y_n)'$ is a vector of responses,
$\bX_n$ is an $n \times p$ fixed design matrix,
$\bm{\beta}=(\beta_1,\dots, \beta_p)'$ is an unknown vector of
parameters, $\bm\varepsilon = (\varepsilon_1, \varepsilon_2,
\dots, \varepsilon_n)'$ is the vector of unobservable random
errors, and the superscript ($'$) denotes the transpose of a
vector or matrix. We do not make any distributional assumption about the errors
except that $\bm\varepsilon$ has a cumulative distribution
function $F(\bm\varepsilon)$ with $E(\bm\varepsilon)=\bm0$, and
$E(\bm\varepsilon \bm\varepsilon') = \sigma^2\bm{I}$, where
$\sigma^2$  is finite.

For $n>p$ the classical estimator of $\bm\beta$ by minimizing the least square function and is
given by
\[
 \widehat\bbeta^{\textrm{LSE}}= ({\bX_n}'{\bX_n})^{-1}{\bX_n' \bY}.
\]

However, since we are dealing with a high-dimensional situation, i.e. $n<p$ so  $(\bX'\bX)^{-1}$ will not exist and thus no solution. However, one can employ the generalized inverse to revert the problem. In the current set-up we are assuming that the model is sparse so it is desirable to use penalized likelihood method to obtain a meaningful solution as was briefly discussed in our Introduction section.
Penalty estimators are a class of estimators in the least penalized
squares family of estimators, see
\cite{Ahmed2014a}. This method involves penalizing the regression coefficients, and shrinking a subset of them to
zero.
In other words, the penalized procedure produces a submodel
and subsequently estimates the submodel parameters. Several
penalty estimators have been proposed in the literature for linear
and generalized linear models. In this section, we consider the LASSO and the ALASSO. By shrinking some
regression coefficients to zero, these methods select parameters
and estimation simultaneously.
\cite{Frank1993} introduced bridge regression, a
generalized version of penalty (or absolute penalty type)
estimators. For a given penalty function $\pi(\cdot)$ and
regularization parameter $\lambda$, the general form can be
written as
\begin{equation*}
S(\bbeta)=(\bY -\bX_n\bbeta)'(\bY -\bX_n\bbeta) + \lambda
\pi(\beta),
\end{equation*}
where the penalty function is of the form
\begin{equation}\label{eq:general:penalizedLS}
\pi(\beta)=\sum_{j=1}^m|\beta_j|^\gamma, \ \gamma > 0.
\end{equation}
The penalty function in (\ref{eq:general:penalizedLS}) bounds the
$L_\gamma$ norm of the parameters in the given model as
$\sum_{j=1}^m|\beta_j|^\gamma \le t$, where $t$ is the tuning
parameter that controls the amount of shrinkage. We see that for
$\gamma=2$, we obtain ridge estimates which are obtained by
minimizing the penalized residual sum of squares
\begin{equation}\label{eq:ridge:solution}
 \widehat\bbeta^{\textrm{Ridge}} = \mbox{arg}\min_{\beta}\left\{\sum_{i=1}^n (y_i
- \beta_0 - \sum_{j=1}^{p}x_{ij}\beta_j)^2 + \lambda
\sum_{j=1}^{p}\beta_j^2 \right\},
\end{equation}
where $\lambda$ is the tuning parameter which controls the amount
of shrinkage.
\cite{Frank1993} did not solve for the bridge regression
estimators for any $\gamma >0.$ Interestingly, for $\gamma < 2$,
it shrinks the coefficient towards zero, and depending on the
value of $\lambda,$ it sets some of them to be exactly zero. Thus,
the procedure combines variable selection and shrinking of the
coefficients of penalized regression. \cite{gao-et-al2016} suggested weighted ridge estimator for high dimensional setting, and investigated the advantages of post selection positive part of shrinkage estimators both theoretically and numerically.

An important member of the penalized least squares family is the
$L_1$ penalized least squares estimator, which is obtained when
$\gamma =1$, and is called LASSO.
\subsection{LASSO}

The least absolute shrinkage and
selection operator was proposed by \cite{Tibshirani1996}, which
performs variable selection and parameter estimation
simultaneously. LASSO is closely related with ridge regression.
LASSO solutions are similarly defined by replacing the squared
penalty $\sum_{j=1}^{p} \beta_j^2$ in the ridge solution
(\ref{eq:ridge:solution}) with the absolute penalty
$\sum_{j=1}^{p}|\beta_j|$ in the LASSO,
\begin{equation}\label{eq:LASSO:solution}
  \widehat\bbeta^{\textrm{LASSO}} = \mbox{arg}\min_{\beta}\left\{\sum_{i=1}^n (y_i
- \beta_0 - \sum_{j=1}^{p}x_{ij}\beta_j )^2 + \lambda
\sum_{j=1}^{p}|\beta_j| \right\}.
\end{equation}

Although the change apparently looks subtle, the absolute penalty
term  made it impossible to have an analytic solution for the
LASSO. Originally, LASSO solutions were obtained via quadratic
programming. Later, \cite{Efron2004} proposed Least Angle
Regression (LAR), a type of stepwise regression, with which the
LASSO estimates can be obtained at the same computational cost as
that of an ordinary least squares estimation. Further, the LASSO
estimator remains numerically feasible for dimensions of $p$ that
are much higher than the sample size $n$.

\subsection{Adaptive LASSO}

\cite{Zou2006} modified the LASSO penalty by using adaptive
weights on $L_1$ penalties on the regression coefficients. Such a
modified method was referred to as ALASSO \index{Adaptive
LASSO}. It has been shown theoretically that the ALASSO
estimator is able to identify the true model consistently, and the
resulting estimator is as efficient as the oracle\index{Oracle
property}.

The ALASSO of
$\widehat\bbeta^{\textrm{ALASSO}}$ are obtained by
\begin{equation}\label{eq:adALASSO:ch4}
 \widehat{\bm\beta}^{\textrm{ALASSO}} = \textrm{arg}\min_{\beta} \left\{
\sum_{i=1}^{n} (y_i - \beta_0 - \sum_{j=1}^{p}x_{ij}\beta_j)^2 +
\lambda \sum_{j=1}^{p} \widehat{w}_j |\beta_j| \right\},
\end{equation}
where the weight function is
\[
 \widehat{w}_j = \frac{1}{|\widehat{\beta}^*_j|^\gamma}; \quad \gamma>0,
\]
and $\widehat{\beta}_j^*$ is a root-n consistent estimator of $\beta$.
Equation~(\ref{eq:adALASSO:ch4}) is a ``convex optimization problem
and its global minimizer can be efficiently solved''
\citep{Zou2006}.


The main objective of this research article is to improve the estimation accuracy of the active set of the regression parameters by combining an over-fitted model estimators with an under-fitted one. For this purpose, we follow the methodology of double shrunken estimator of \cite{YuzbasiArashi2016}. As stated earlier, the LASSO produce an over-fitted model as compared with ALASSO and other variable selection methods. The LASSO strategy retains some regression coefficients with weak effects and as well as some with weak effects in the resulted model. On the other hand, aggressive variable selection strategies may force moderate and effects coefficients towards zero, resulting in under-fitted models with a fewer variable of strong effect. The idea here is to combine estimators from an under-fitted model with an over-fitted model using a non-linear shrinkage technique incorporating a measurable bounded function.

\section{Double Shrunken Estimators}
In this section, we show how to shrink more the addressed penalized estimators, in the combination of two submodels produced by two distinct variable selection techniques. Similar to \cite{AhmedYuzbasi2016}, the idea is to work with a sparse model that will be all the predictors included and then apply two variable selection methods with high and low penalties, respectively. Finally, we combine the estimates from two models to improve post estimation and prediction performances, respectively and incorporate the concept of double shrunken of \cite{YuzbasiArashi2016}.

\subsection{Working Model}

Consider the following high dimensional sparse regression model with strong and weak-to-moderate signals
\begin{equation}\label{eq:ch1:full:model}
\bY=\bX_n \bm\beta+ \bm \varepsilon, \quad p > n
\end{equation}
Suppose we can divide the index set $\{1,\cdots, p\}$ into
three disjoint subsets:  $S_{1}$, $S_{2}$ and $S_{3}$. In particular,
$S_{1}$ includes indexes of nonzero
$\beta_i$'s which are large and comfortably detectable. The set $S_{2}$, being the intermediate,
  includes indexes of those  nonzero $\beta_j$ with  weak-to-moderate but nonzero effects.
 By the assumption of sparsity $S_{3}$ includes indexes with only zero coefficients and can be easily discarded by exiting variable selection methods.  Thus, $S_{1}$ and $S_{3}$ are able to be retained and discarded by using existing variable selection techniques, respectively. However, it is possible that
 the $S_2$ may be covertly included either in $S_{2}$ or $S_{3}$ depending on existing LASSO-type methods. For the case when $S_2$ may not be separated from $S_{3}$, some work has been done in this area, see \cite{Zhang2014} and others. \cite{hansen2015} has showed using
  simulation studies that such a LASSO estimate
often performs worse than the post selection least square estimate.
To improve the prediction error of  a LASSO-type variable selection approach,
some (modified) post least squares estimators are studied in \cite{bellCher2009} and \cite{liuYu2013}.

However, we are interested in cases when covariates in $S_{1}$  are  kept in the model, and some or all covariates in $S_{2}$ are also included in  $S_{1}$, which may or may not be useful for prediction purposes. It is possible that one variable selection strategies may produce an over-fitted model, that is retaining predictors from  $S_{1}$ and $S_{2}$. On the other hand, other methods may produce an under-fitted model keeping only predictors from  $S_{1}$. Thus, the predictors in $S_{2}$ should be subject to further scrutiny to improve the prediction error.

We partition the design matrix
such that $\bX=(\bX_{S_1}| \bX_{S_2}| \bX_{S_3})$,
Further, $\bX_{n1}$ is $n  \times p_1$,
$\bX_{n2}$ is $n \times p_2$, and $\bX_{n3}$ is $n \times p_3$ submatrix of predictors, respectively; and $p=p_1+p_2+p_3$.  Here we make the usual assumption that $p_1 \le p_2 <n$ and $p_3>n$.

Thus, our working model is rewritten as:
\begin{equation}\label{eq:ch1:full:model}
\bY=\bX_{n1} \bm\beta_1+ \bX_{n2} \bm\beta_2+ \bX_{n3} \bm\beta_3+\bm \varepsilon, \quad p > n, \ \ p_1+p_2 <n.
\end{equation}

\subsection{Overfitted Model}

We apply a variable selection method which keeps both strong and weak-moderate signals as follows:

\begin{equation}\label{eq:ch1:full:model1}
\bY=\bX_{n1} \bm\beta_1+ \bX_{n2} \bm\beta_2+ \varepsilon, \quad p_1 \le p_2 <n.
\end{equation}
Recall, the LASSO strategy which usually eliminates the sparse signals and retains weak-moderate and strong signals in the resulting model, and may be considered as an overfitted Model

\subsection{Underfitted Model}

Now, we apply a variable selection method which keeps only strong signals and eliminates all other signals in the resulting model. Thus, we have

\begin{equation}\label{eq:ch1:full:model2}
\bY=\bX_{n1} \bm\beta_1+ \varepsilon, \quad p_1 <n.
\end{equation}
One can use ALASSO strategy which usually retain the strong signals and may produce a lower dimensional model as compared with LASSO.  This model may bay termed as an underfitted Model.

We are interested in estimating $\boldsymbol{\beta_1}$ when $\bm\beta_2$ may be a null vector, but we are not sure. We suggest Stein-type shrinkage strategy for estimating $\bm\beta_1$ under this real situation. In  essence we would like to combine estimates of the overfitted with the estimates of underfitted models to improve the efficiency of an underfitted model.

\subsection{Double Shrinking}

\cite{AhmedYuzbasi2016} defined a shrinkage estimator of $\boldsymbol{\beta
}_1$ by combining overfitted model estimate $\boldsymbol{\widehat{\beta}}_1^{\rm OF}$ with the underfitted $\boldsymbol{\widehat{\beta}}_1^{\rm UF}$ as
\begin{equation}
\boldsymbol{\widehat{\beta}}_1^{S}=\boldsymbol{\widehat{\beta}}_1^{\rm UF}+\left( \boldsymbol{\widehat{\beta}}_1^{\rm OF}-\boldsymbol{%
\widehat{\beta}}_1^{\rm UF}\right) \left( 1-(p_{2}-2)W_{n}^{-1}\right) \text{, }p_{2}\geq 3,
\end{equation}
where, the weight function $W_{n}$  is defined by
\begin{equation*}
W_{n}=\frac{n}{\widehat{\sigma}^{2}} (\boldsymbol{\widehat{%
\beta}}_{2}^{LSE})'(\bX_{S_2}'\boldsymbol{M}_{1}%
\bX_{S_2})\boldsymbol{\widehat{\beta}}_{2}^{LSE},
\end{equation*}
and $\boldsymbol{M}_{1}=\boldsymbol{I}_{n}-\bX_{S_1}\left(
\bX_{S_1}'\bX_{S_1}\right) ^{-1}\bX%
_{S_1}'$, $\boldsymbol{\widehat{\beta}}_{2}^{LSE}=\left( \bX%
_{S_2}'\boldsymbol{M}_{1}\bX_{S_2}\right) ^{-1}\bX%
_{S_2}'\boldsymbol{M}_{1}\bY$
and
\begin{equation*}
\widehat{\sigma}^{2}=\frac{1}{n-1}(\bY-\bX_{S_1}\boldsymbol{%
\widehat{\beta}}_1^{\rm UF})'(\bY-\bX_{S_1}\boldsymbol{%
\widehat{\beta}}_1^{\rm UF}).
\end{equation*}%
The $\boldsymbol{\widehat{\beta}}_1^{\rm UF}$ is the ALASSO estimator and $\boldsymbol{\widehat{\beta}}_1^{\rm OF}$ is the LASSO estimator.

Here, under the concept of double shrinking of \cite{YuzbasiArashi2016}, we define a family of double shrunken estimators
\begin{eqnarray}
\boldsymbol{\widehat{\beta}}_1^{\rm FS}&=&\boldsymbol{\widehat{\beta}}_1^{\rm OF}-\frac{(p_{2}-2)r(W_n)}{W_{n}}\left( \boldsymbol{\widehat{\beta}}_1^{\rm OF}-\boldsymbol{%
\widehat{\beta}}_1^{\rm UF}\right)\cr
&=&\boldsymbol{\widehat{\beta}}_1^{\rm UF}+\left( \boldsymbol{\widehat{\beta}}_1^{\rm OF}-\boldsymbol{%
\widehat{\beta}}_1^{\rm UF}\right) \left( 1-\frac{(p_{2}-2)r(W_n)}{W_{n}}\right) \text{, }p_{2}\geq 3,
\end{eqnarray}
where $r(x)$ is a continuous, bounded and differentiable function of $x$.

For $r(x)=1$, we get the result of \citet{AhmedYuzbasi2016}.

In the sprit of \citet{AlamThompson1969}, we consider the function $r(x)=1/(1+x^{-1})$ to get
\begin{eqnarray}
\boldsymbol{\widehat{\beta}}_1^{\rm FS1}&=&\boldsymbol{\widehat{\beta}}_1^{\rm UF}+\left( \boldsymbol{\widehat{\beta}}_1^{\rm OF}-\boldsymbol{%
\widehat{\beta}}_1^{\rm UF}\right) \left( 1-\frac{(p_{2}-2)}{1+W_{n}}\right) \text{, }p_{2}\geq 3,
\end{eqnarray}
Further, by the virtue of Gaussian kernel, we consider the function $r(x)=\exp(-x^2)$ to get
\begin{eqnarray}
\boldsymbol{\widehat{\beta}}_1^{\rm FS2}&=&\boldsymbol{\widehat{\beta}}_1^{\rm UF}+\left( \boldsymbol{\widehat{\beta}}_1^{\rm OF}-\boldsymbol{%
\widehat{\beta}}_1^{\rm UF}\right) \left( 1-\frac{(p_{2}-2)\exp(-W_n^2)}{W_{n}}\right) \text{, }p_{2}\geq 3,
\end{eqnarray}
Lastly, we propose to use $r(x)=\arctan(x)$, which yields the following superior estimator
\begin{eqnarray}
\boldsymbol{\widehat{\beta}}_1^{\rm FS3}&=&\boldsymbol{\widehat{\beta}}_1^{\rm UF}+\left( \boldsymbol{\widehat{\beta}}_1^{\rm OF}-\boldsymbol{%
\widehat{\beta}}_1^{\rm UF}\right) \left( 1-\frac{(p_{2}-2)\arctan(W_n)}{W_{n}}\right) \text{, }p_{2}\geq 3,
\end{eqnarray}
In the forthcoming section we will be analyzing the performance of $\boldsymbol{\widehat{\beta}}_1^{\rm FS1}$ and $\boldsymbol{\widehat{\beta}}_1^{\rm FS3}$ and compare with the superior estimator of \cite{AhmedYuzbasi2016}, i.e., the PS estimator. In the conclusions, we will discuss about the usage of $\boldsymbol{\widehat{\beta}}_1^{\rm FS2}$.

\section{Theoretical Considerations}
In this section, we develop some properties of the proposed estimators. Because of the complexity, we only consider orthonormal design.
Note that in general, the LASSO is not an oracle procedure and is not consistent, whereas the ALASSO has oracle properties. Honestly, our result, is restrictive.

For our purpose, we assume $n^{-1}\bX'\bX=\bI_p$. Under the specified partitioning, $n^{-1}\bX_{S_i}'\bX_{S_i}=\bI_{p_i}$ and $\bX_{S_i}'\bX_{S_j}=\boldsymbol 0$, for $i\neq j=1,2,3$. Simply, $\boldsymbol{M}_{1}=\bI_n-\bX_{S_1}\bX_{S_1}'$, $\bX_{S_2}'\boldsymbol{M}_1\bX_{S_2}=n\bI_{p_2}$, and
\begin{equation}
\boldsymbol{\widehat{\beta}}_{2}^{LSE}=n^{-1}\bX_{S_2}'\bY,\quad
W_n=\frac{1}{\widehat{\sigma}^{2}} (\boldsymbol{\widehat{%
\beta}}_{2}^{LSE})'\boldsymbol{\widehat{\beta}}_{2}^{LSE}.
\end{equation}
Further, we have
\begin{eqnarray}
\hat{\boldsymbol\beta}_{1}^{\rm OF}&=&\left({\rm sgn}(\hat\beta_j^{\rm LSE})\left(|\hat\beta_j^{\rm LSE}|-\frac{\lambda}{2}\right)^+,j=1,\ldots,p_1\right)'\cr
\hat{\boldsymbol\beta}_{1}^{\rm UF}&=&\left({\rm sgn}(\hat\beta_j^{\rm LSE})\left(|\hat\beta_j^{\rm LSE}|-\frac{\lambda}{2|\hat\beta_j^{\rm LSE}|}\right)^+,j=1,\ldots,p_1\right)'.
\end{eqnarray}
For the true parameter value, $|\hat\beta_{1j}|>\lambda/2$, for $j=1,\ldots,p_o<p_1$, where $p_o$ is the true parameter value for the active set $\{j:\beta_j\neq0,j=1,\ldots,p_1\}$. Then, it is easy to see
\begin{equation*}
\boldsymbol{\widehat{\beta}}_1^{\rm OF}-\boldsymbol{\widehat{\beta}}_1^{\rm UF}=
\left(\frac{\lambda}{2}{\rm sgn}(\hat\beta_j^{\rm LSE})\left(\frac{1}{|\hat\beta_j^{\rm LSE}|}-1\right),j=1,\ldots,p_o\right)'.
\end{equation*}
Therefore, we can obtain the following bound
\begin{equation}\label{diff}
\mathcal{\bD}=\boldsymbol{\widehat{\beta}}_1^{\rm OF}-\boldsymbol{\widehat{\beta}}_1^{\rm UF}<{\rm sgn}(\hat\beta_j^{\rm LSE})(1-\frac{\lambda}{2})
\end{equation}
if $\lambda$ is suitably chosen such that $\lambda<2$.

Define the risk function of any estimator $\hat\bbeta_1$ of true parameter $\bbeta_1$ by ${\rm R}(\bbeta_1;\hat\bbeta_1)=\mathbb{E}(\hat\bbeta_1-\bbeta_1)'(\hat\bbeta_1-\bbeta_1)$. Note, here we have $j=1,\ldots,p_o$.

Then, under the orthonormal assumption, we have
\begin{eqnarray}
{\rm R}(\bbeta;\boldsymbol{\widehat{\beta}}_1^{\rm FS})-{\rm R}(\bbeta;\boldsymbol{\widehat{\beta}}_1^{\rm OF})&=&
(p_2-2)^2\mathbb{E}\left[\frac{r^2(W_n)}{W_n^2}\mathcal{\bD}'\mathcal{\bD}\right]\cr
&&-2(p_2-2)\mathbb{E}\left[\frac{r(W_n)}{W_n}(\boldsymbol{\widehat{\beta}}_1^{\rm OF}-\bbeta_1)'\mathcal{\bD}\right]\cr
&<&(1-\frac{\lambda}{2})^2(p_2-2)^2\mathbb{E}\left[\frac{r^2(W_n)}{W_n^2}\sum_{j=1}^{p_o}{\rm sgn}(\hat\beta_j^{\rm LSE})\right]\cr
&&-2(p_2-2)(1-\frac{\lambda}{2})\mathbb{E}\left[\frac{r(W_n)}{W_n}(\boldsymbol{\widehat{\beta}}_1^{\rm OF}-\bbeta_1)'{\rm sgn}(\hat{\boldsymbol\beta}^{\rm LSE})\right]
\end{eqnarray}
As $n\to\infty$, $\boldsymbol{\widehat{\beta}}_1^{\rm OF}\overset{P}{\to}\bbeta_1$. Hence, for sufficiently large samples size $n$, ${\rm R}(\bbeta;\boldsymbol{\widehat{\beta}}_1^{\rm FS})<{\rm R}(\bbeta;\boldsymbol{\widehat{\beta}}_1^{\rm OF})$, i.e., the proposed $\boldsymbol{\widehat{\beta}}_1^{\rm FS}$ outperforms $\boldsymbol{\widehat{\beta}}_1^{\rm OF}$ ($\boldsymbol{\widehat{\beta}}_1^{\rm FS}\succ\boldsymbol{\widehat{\beta}}_1^{\rm OF}$) as soon as $\sum_{j=1}^{p_o}{\rm sgn}(\hat\beta_j^{\rm LSE})<0$, under a probabilistic sense. This scenario is independent of the choice of $r(\cdot)$ and hence, all the shrinkage estimators outperform the over fitted model. Similar conclusion can be discovered for the under fitted model, with a slightly different condition.

In general, $\boldsymbol{\widehat{\beta}}_1^{\rm FS}\succ\boldsymbol{\widehat{\beta}}_1^{\rm OF}$ iff for all $r(\cdot)$, we have
\begin{eqnarray}\label{eq45}
\mathbb{E}\left[\frac{r(W_n)}{W_n}\left\{(p_2-2)\frac{r(W_n)}{W_n}\mathcal{\bD}-2(\boldsymbol{\widehat{\beta}}_1^{\rm OF}-\bbeta_1)\right\}'\mathcal{\bD}\right]<0
\end{eqnarray}
Let 
\begin{equation}
\alpha=\frac{(\boldsymbol{\widehat{\beta}}_1^{\rm OF}-\bbeta_1)'\mathcal{\bD}}{\mathcal{\bD}'\mathcal{\bD}}
\end{equation}
Then, $\alpha$ satisfies $\sqrt n(\boldsymbol{\widehat{\beta}}_1^{\rm OF}-\bbeta_1)=\alpha\sqrt{n}(\boldsymbol{\widehat{\beta}}_1^{\rm OF}-\boldsymbol{\widehat{\beta}}_1^{\rm UF})=\alpha\left[\sqrt n(\boldsymbol{\widehat{\beta}}_1^{\rm OF}-\bbeta_1)-
\sqrt n(\boldsymbol{\widehat{\beta}}_1^{\rm UF}-\bbeta_1)\right]$. Let $\lambda=o(\sqrt n)$ and $\lambda n^{(\gamma-1)/2}\to\infty$. Using Theorem 2 of \cite{Zou2006}, $\sqrt n(\boldsymbol{\widehat{\beta}}_1^{\rm UF}-\bbeta_1)\overset{P}{\to}0$. Consequently $\alpha\to1$. Now, we are ready to find the bound on $r(W_n)/W_n$.

Suppose $r(\cdot)>0$ and is concave. Then, using Lemma 1 of \cite{Casella1990}, $r(W_n)/W_n$ is non-increasing. Hence, by \eqref{eq45}, $\boldsymbol{\widehat{\beta}}_1^{\rm FS}\succ\boldsymbol{\widehat{\beta}}_1^{\rm OF}$ if for every $r$ function we have
\begin{eqnarray}
\frac{r(W_n)}{W_n}&<&\frac{2}{p_2-2}\frac{(\boldsymbol{\widehat{\beta}}_1^{\rm OF}-\bbeta_1)'\mathcal{\bD}}{\mathcal{\bD}'\mathcal{\bD}}\to\frac{2}{p_2-2}.
\end{eqnarray}

\section{Monte Carlo Simulation}

We consider a Monte Carlo simulation, and simulate the response from the following model:
\begin{equation}
y_{i}=x_{1i}\beta _{1}+x_{2i}\beta _{2}+...+x_{pi}\beta _{p}+\varepsilon _{i\text{ }},\ i=1,2,...,n,  \label{sim.mod}
\end{equation}%
where $\varepsilon _{i}$ are i.i.d. $N\left( 0,1\right)$ and ${x}_{ij}=(\xi_{(ij)}^1)^2+\xi_{(ij)}^2$ with $\xi_{(ij)}^1\sim N\left( 0,1\right)$ and $\xi_{(ij)}^2\sim N\left( 0,1\right)$ for all $i=1,2,...,n$, $j=1,2,...,p$.

We consider the regression coefficients are set
$\bbeta=\left( \bbeta_{1}',\bbeta_{2}',\bbeta_{3}'\right)' =\left( \bm\Lambda'_{p_1},\bm\Delta'_{p_2},\bm{0}_{p_3}'\right)'$, where, $\bm\Lambda_{p_1}$, $\bm\Delta_{p_2}$ and $\bm{0}_{p_3}$ mean the vectors of $\Lambda$, $\Delta$ and 0 with dimensions $p_1$, $p_2$ and $p_3$, respectively. If $\Delta=0$, then it indicates that the null hypothesis is true. On the other hand, the larger values $\Delta$ indicate the degree of violation of null hypothesis.

In this simulation setting, we simulated 250 data sets consisting of $n=150$, $\Lambda=1,2$, $p_1=4$, $p_2=4,8,16$ and $p_3=200,400,800$.

\begin{figure}[!htbp]
\centering
\includegraphics[width=14cm,height=12cm]{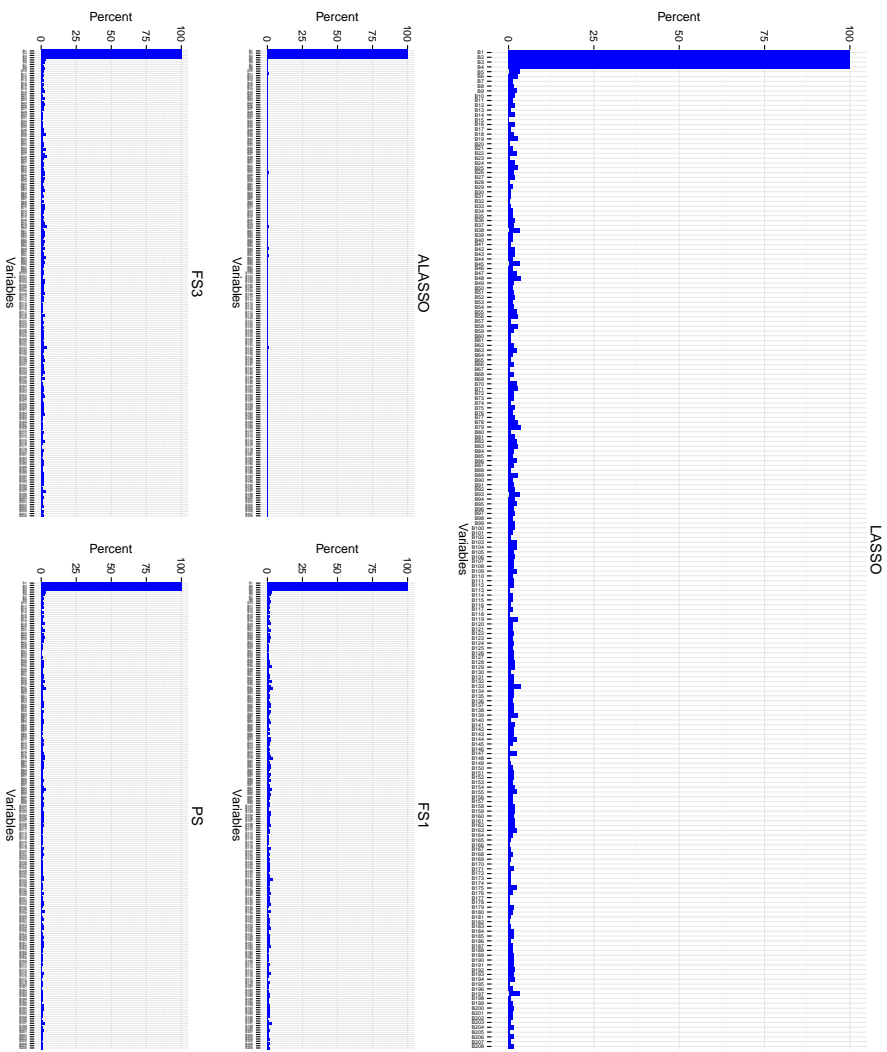}
\caption{The percentage of times each predictor was selected when $\Delta=0$ and $(p_2,p_3)=(4,200)$
 \label{Fig:delta0}}
\end{figure}

\begin{figure}[!htbp]
\centering
\includegraphics[width=14cm,height=12cm]{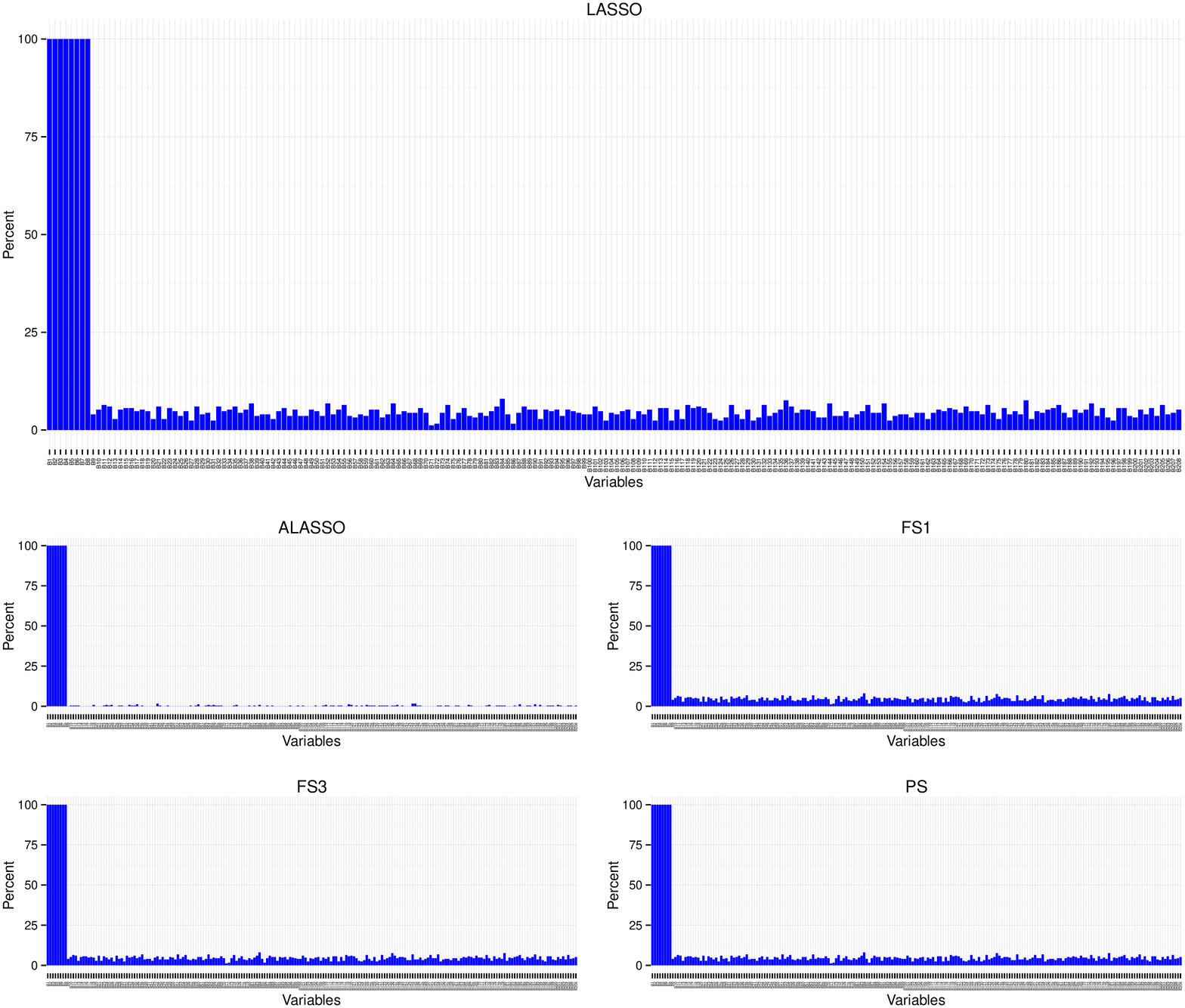}
\caption{The percentage of times each predictor was selected when $\Delta=0.8$ and $(p_2,p_3)=(4,200)$
 \label{Fig:delta08}}
\end{figure}

The performance of an estimator is evaluated by using relative mean squared error (RMSE) criterion. The RMSE of an estimator $\boldsymbol{\beta }_1^{\ast }$ with respect to $\boldsymbol{\widehat{\beta }}_1^{\rm OF}$ is defined as follows%
\begin{equation}
\label{RE}
\textnormal{RMSE}\left( \boldsymbol{{\beta }}_1%
^{\ast } \right) =\frac{ \rm MSE \left( \boldsymbol{\widehat{\beta }}_1%
^{\rm OF} \right) }{\rm MSE \left( \boldsymbol{{\beta }}_1%
^{\ast} \right)},
\end{equation}
where $\boldsymbol{\beta }_{1}^{\ast}$ is one of the listed estimators. If the RMSE of an estimator is larger than one, it indicates that it is superior to $\boldsymbol{\widehat{\beta }}_1^{\rm OF}$. The results of simulated RMSE of the listed estimators are reported in Tables \ref{Tab1} -- \ref{Tab6} and Figures \ref{Fig:delta0} and \ref{Fig:delta08}. We also report the TP (the number of true positives) and the FP (the number of false positives) in Table \ref{Tab:TFP} only for $(p_2,p_3)=(4,200)$.

According to the simulation results, the performance of under-fitted estimator ALASSO is the best since it is based on true model, and the FS3 performs better than both FS1 and PS when $\Delta=0$. On the other hand, the RMSE of the ALASSO decreases and approaches to zero while the all others approach to one when we increase the magnitude of weak signals.

In Figure \ref{Fig:delta0}, if $\Delta=0$, then both LASSO and ALASSO methods always select strong covariates, while ALASSO select less weak signals than LASSO. Contrary to this, if we increase the magnitude of weak signals, say $\Delta=0.8$, then we observe that LASSO is more efficient than ALASSO for selecting those signals, see the Figure \ref{Fig:delta08}. For both case, our suggest methods again select all strong signals, while they select more variables than ALASSO when the weak signals are getting stronger.

\begin{table}[!htbp]
\caption{The numbers and the percentages of TP and FP of estimators when $(p_2,p_3)=(4,200)$}
\centering
\begin{tabular}{crrrrrrrrrr}
  \hline  \hline

  $\Delta$ & 0.000 & 0.200 & 0.400 & 0.600 & 0.800 & 0.000 & 0.200 & 0.400 & 0.600 & 0.800 \\

  \hline  \hline

  && \multicolumn{4}{c}{$\#$ the number of TP}&&\multicolumn{4}{c}{$\#$ the number of FP} \\
\cmidrule(lr){2-6} \cmidrule(lr){7-11}
   LASSO & 4.00 & 4.00 & 4.00 & 4.00 & 4.00 & 3.06 & 6.21 & 12.98 & 15.56 & 16.62 \\
  ALASSO & 4.00 & 4.00 & 4.00 & 4.00 & 4.00 & 0.16 & 0.95 & 4.03 & 5.28 & 5.18 \\
  FS1 & 4.00 & 4.00 & 4.00 & 4.00 & 4.00 & 3.06 & 6.21 & 12.98 & 15.56 & 16.62 \\
  FS3 & 4.00 & 4.00 & 4.00 & 4.00 & 4.00 & 3.06 & 6.21 & 12.98 & 15.56 & 16.62 \\
  PS & 4.00 & 4.00 & 4.00 & 4.00 & 4.00 & 2.06 & 6.21 & 12.98 & 15.56 & 16.62 \\
   \hline

     && \multicolumn{4}{c}{$\#$ the percentages of TP}&&\multicolumn{4}{c}{$\#$ the percentages of FP} \\
\cmidrule(lr){2-6} \cmidrule(lr){7-11}

  LASSO & 100.00 & 100.00 & 100.00 & 100.00 & 100.00 & 0.76 & 1.54 & 3.21 & 3.85 & 4.11 \\
  ALASSO & 100.00 & 100.00 & 100.00 & 100.00 & 100.00 & 0.04 & 0.24 & 1.00 & 1.31 & 1.28 \\
  FS1 & 100.00 & 100.00 & 100.00 & 100.00 & 100.00 & 0.76 & 1.54 & 3.21 & 3.85 & 4.11 \\
  FS3 & 100.00 & 100.00 & 100.00 & 100.00 & 100.00 & 0.76 & 1.54 & 3.21 & 3.85 & 4.11 \\
  PS & 100.00 & 100.00 & 100.00 & 100.00 & 100.00 & 0.51 & 1.54 & 3.21 & 3.85 & 4.11 \\
      \hline
\end{tabular}
\label{Tab:TFP}
\end{table}

Table \ref{Tab:TFP} shows the numbers and the percentages of TP and FP of the listed estimators when $(p_2,p_3)=(4,200)$. According to this table, all listed methods select all strong covariates for each values of $\Delta$, whereas ALASSO is the best for FP, which is indicated by the smallest ratio of FP.

\begin{table}[!htbp]
\caption{RMSE of estimators when $p_3 = 200$ and $\Lambda=1$}
\label{Tab1}
\centering
\begin{tabular}{ccrrrrr}
  \hline
$p_2$& $\Delta$ & ALASSO & FS1 & FS3 & PS \\
  \hline
4 &  0.000  & 1.717 & 1.325 & 1.507 & 1.399 \\
  &0.200  & 1.324 & 1.040 & 1.065 & 1.043 \\
  &0.400 &0.670 & 1.004 & 1.006 & 1.004 \\
  &0.600 & 0.324 & 0.994 & 0.991 & 0.994 \\
  &0.800 &0.189 & 0.992 & 0.988 & 0.992 \\ \hline
8&  0.000 &  1.727 & 1.547 & 1.689 & 1.574 \\
  &0.200 & 1.238 & 1.051 & 1.080 & 1.052 \\
  &0.400 & 0.516 & 0.997 & 0.995 & 0.997 \\
  &0.600 &0.235 & 0.985 & 0.976 & 0.985 \\
  &0.800 &0.142 & 0.984 & 0.974 & 0.984 \\ \hline
16&  0.000 &1.749 & 1.677 & 1.888 & 1.653 \\
  &0.200 & 1.104 & 1.050 & 1.076 & 1.050 \\
  &0.400 &0.458 & 0.993 & 0.988 & 0.993 \\
  &0.600 & 0.200 & 0.979 & 0.967 & 0.979 \\
  &0.800 &  0.113 & 0.979 & 0.967 & 0.979 \\
   \hline
\end{tabular}
\end{table}

\begin{table}[!htbp]
\caption{RMSE of estimators when $p_3 = 400$ and $\Lambda=1$}
\label{Tab2}
\centering
\begin{tabular}{ccrrrrr}
  \hline
$p_2$& $\Delta$ &  ALASSO & FS1 & FS3 & PS \\
  \hline
4& 0.000 &  1.923 & 1.401 & 1.653 & 1.496 \\
  &0.200 & 1.483 & 1.046 & 1.073 & 1.048 \\
  &0.400 &0.827 & 1.008 & 1.012 & 1.008 \\
  &0.600 &0.382 & 0.996 & 0.994 & 0.996 \\
  &0.800 &  0.227 & 0.994 & 0.990 & 0.994 \\ \hline
8&  0.000 &  1.920 & 1.674 & 1.939 & 1.703 \\
  &0.200 &  1.332 & 1.061 & 1.097 & 1.063 \\
  &0.400 & 0.637 & 1.003 & 1.004 & 1.003 \\
  &0.600 &  0.295 & 0.989 & 0.983 & 0.989 \\
  &0.800 &0.166 & 0.986 & 0.978 & 0.986 \\ \hline
16& 0.000 & 1.927 & 1.789 & 2.078 & 1.776 \\
  &0.200 &  1.213 & 1.056 & 1.087 & 1.057 \\
  &0.400 &  0.634 & 1.002 & 1.002 & 1.002 \\
  &0.600 & 0.287 & 0.988 & 0.980 & 0.988 \\
  &0.800 &  0.158 & 0.984 & 0.975 & 0.984 \\
   \hline
\end{tabular}
\end{table}

\begin{table}[!htbp]
\caption{RMSE of estimators when $p_3 = 800$ and $\Lambda=1$}
\label{Tab3}
\centering
\begin{tabular}{ccrrrrrrrr}
  \hline
$p_2$& $\Delta$ &  ALASSO & FS1 & FS3 & PS \\
  \hline
4&  0.000 & 2.131 & 1.455 & 1.740 & 1.579 \\
  &0.200 &  1.587 & 1.052 & 1.083 & 1.055 \\
  &0.400 &  0.977 & 1.011 & 1.018 & 1.011 \\
  &0.600 &0.468 & 0.998 & 0.997 & 0.998 \\
  &0.800 &  0.265 & 0.995 & 0.992 & 0.995 \\ \hline
8&  0.000 & 2.255 & 1.777 & 1.955 & 1.874 \\
  &0.200 &  1.472 & 1.069 & 1.110 & 1.071 \\
  &0.400 &  0.822 & 1.009 & 1.014 & 1.009 \\
  &0.600 & 0.382 & 0.994 & 0.990 & 0.994 \\
  &0.800 &0.212 & 0.990 & 0.984 & 0.990 \\ \hline
16&  0.000 & 2.175 & 2.039 & 2.406 & 1.984 \\
  &0.200 &  1.325 & 1.065 & 1.103 & 1.066 \\
  &0.400 & 0.842 & 1.009 & 1.013 & 1.009 \\
  &0.600 &  0.468 & 0.996 & 0.994 & 0.996 \\
  &0.800 &  0.261 & 0.991 & 0.985 & 0.991 \\
   \hline
\end{tabular}
\end{table}

\begin{table}[!htbp]
\caption{RMSE of estimators when $p_3 = 200$ and $\Lambda=2$}
\label{Tab4}
\centering
\begin{tabular}{ccrrrr}
  \hline
$p_2$& $\Delta$  & ALASSO & FS1 & FS3 & PS \\
  \hline
4&  0.000 & 1.799 & 1.356 & 1.596 & 1.436 \\
&0.200 & 1.249 & 1.032 & 1.051 & 1.034 \\
  &0.400 & 0.690 & 1.003 & 1.005 & 1.003 \\
  &0.800 & 0.193 & 0.993 & 0.990 & 0.993 \\
  &1.200 & 0.085 & 0.993 & 0.990 & 0.993 \\
  &1.600 & 0.048 & 0.994 & 0.991 & 0.994 \\ \hline
 8& 0.000 & 1.758 & 1.531 & 1.752 & 1.557 \\
  & 0.200 & 1.081 & 1.031 & 1.048 & 1.032 \\
  &0.400 & 0.518 & 0.994 & 0.990 & 0.994 \\
  &0.800 & 0.134 & 0.984 & 0.974 & 0.984 \\
  &1.200 & 0.060 & 0.986 & 0.978 & 0.986 \\
  &1.600 & 0.034 & 0.988 & 0.981 & 0.988 \\ \hline
16&  0.000 & 1.766 & 1.689 & 2.048 & 1.652 \\
 &0.200 & 1.001 & 1.029 & 1.043 & 1.030 \\
  &0.400 & 0.472 & 0.992 & 0.987 & 0.992 \\
  &0.800 & 0.114 & 0.979 & 0.967 & 0.979 \\
  &1.200 & 0.051 & 0.982 & 0.972 & 0.982 \\
  &1.600 & 0.028 & 0.985 & 0.976 & 0.985 \\
   \hline
\end{tabular}
\end{table}

\begin{table}[!htbp]
\caption{RMSE of estimators when $p_3 = 400$ and $\Lambda=2$}
\label{Tab5}
\centering
\begin{tabular}{ccrrrr}
  \hline
$p_2$& $\Delta$  & ALASSO & FS1 & FS3 & PS \\
  \hline
4&0.000 & 1.934 & 1.427 & 1.734 & 1.547 \\
&0.200 & 1.459 & 1.044 & 1.070 & 1.047 \\
  &0.400 & 0.821 & 1.006 & 1.010 & 1.007 \\
  &0.800 & 0.224 & 0.994 & 0.991 & 0.994 \\
  &1.200 & 0.100 & 0.994 & 0.991 & 0.994 \\
  &1.600 & 0.060 & 0.995 & 0.992 & 0.995 \\ \hline
8 & 0.000 & 2.052 & 1.743 & 2.041 & 1.784 \\
 & 0.200 & 1.247 & 1.043 & 1.068 & 1.044 \\
  &0.400 & 0.639 & 1.002 & 1.002 & 1.002 \\
  &0.800 & 0.166 & 0.987 & 0.979 & 0.987 \\
  &1.200 & 0.073 & 0.988 & 0.981 & 0.988 \\
  &1.600 & 0.043 & 0.990 & 0.984 & 0.989 \\ \hline
 16 &0.000 & 2.022 & 1.961 & 2.518 & 1.885 \\
   &0.200 & 1.080 & 1.034 & 1.052 & 1.035 \\
  &0.400 & 0.618 & 1.000 & 0.999 & 1.000 \\
  &0.800 & 0.159 & 0.984 & 0.975 & 0.984 \\
  &1.200 & 0.071 & 0.985 & 0.977 & 0.985 \\
  &1.600 & 0.040 & 0.987 & 0.980 & 0.987 \\
   \hline
\end{tabular}
\end{table}

\begin{table}[!htbp]
\caption{RMSE of estimators when $p_3 = 800$ and $\Lambda=2$}
\label{Tab6}
\centering
\begin{tabular}{ccrrrr}
  \hline
$p_2$& $\Delta$  & ALASSO & FS1 & FS3 & PS \\
  \hline
4&0.000 & 2.396 & 1.524 & 1.903 & 1.686 \\
&0.200 & 1.552 & 1.048 & 1.078 & 1.051 \\
  &0.400 & 0.943 & 1.009 & 1.014 & 1.009 \\
  &0.800 & 0.258 & 0.996 & 0.993 & 0.996 \\
  &1.200 & 0.121 & 0.995 & 0.992 & 0.995 \\
  &1.600 & 0.067 & 0.995 & 0.993 & 0.995 \\ \hline
 8& 0.000 & 2.231 & 1.789 & 2.145 & 1.853 \\
  & 0.200 & 1.361 & 1.058 & 1.092 & 1.059 \\
  &0.400 & 0.803 & 1.007 & 1.011 & 1.007 \\
  &0.800 & 0.217 & 0.990 & 0.984 & 0.990 \\
  &1.200 & 0.099 & 0.990 & 0.984 & 0.990 \\
  &1.600 & 0.053 & 0.991 & 0.986 & 0.991 \\ \hline
  16&0.000 & 2.203 & 2.093 & 2.677 & 2.000 \\
    &0.200 & 1.250 & 1.053 & 1.082 & 1.053 \\
  &0.400 & 0.875 & 1.009 & 1.014 & 1.009 \\
  &0.800 & 0.254 & 0.990 & 0.984 & 0.990 \\
  &1.200 & 0.107 & 0.989 & 0.983 & 0.989 \\
  &1.600 & 0.062 & 0.990 & 0.984 & 0.990 \\
   \hline
\end{tabular}
\end{table}

\section{Real Data Examples}

In this section, we analyze the performance of double shrinkage estimators dealing with two real data sets. We follow \cite{AhmedYuzbasi2016} strategies for analyzing each data sets. As long as there is no uncertain prior information about $p_1$, $p_2$ and $p_3$, one may use LASSO and ALASSO methods to find important covariates. After that, we may construct our estimation strategies. We also indicate that we draw 1000 bootstrap sample with dimension of the design matrix, and we calculate the prediction error (PE) based on 5 - fold cross validation, and take its average value for each bootstrap sample. To easy comparison, we report relative PE (RPE) of an estimator with respect to LASSO. Thus, a value of RPE $>1$ reflects the superiority of the other methods.

\subsection{Eye Data}

This data set contains gene expression data of mammalian eye tissue samples, \cite{Scheetz-et-al}. The format is a list containing the design matrix which represents the data of $n=120$ rats with $p=200$ gene probes and the response vector with 120 dimensional which represents the expression level of TRIM32 gene. The numbers of selected variables for eye data set are 24 and 11 by LASSO and ALASSO, respectively.

\begin{figure}[!htbp]
\centering
\includegraphics[width=15cm,height=8cm]{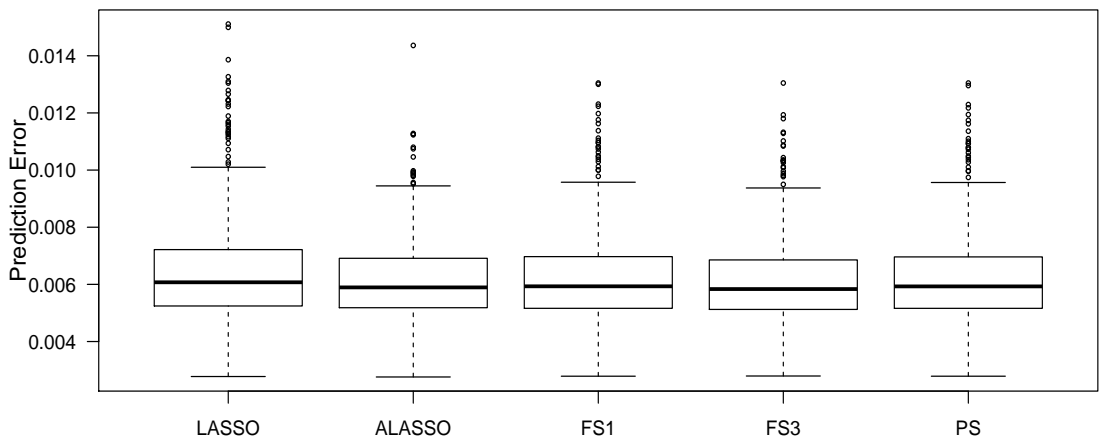}
\caption{The RPEs for Eye data set.
 \label{Fig:eyedata:box}}
\end{figure}

In Figure \ref{Fig:eyedata:box}, we plot the prediction error of each bootstrap replication for listed estimation techniques.
Also, in Table \ref{Tab:eyedata}, we report the RPEs of estimators. It can be seen that the performance of FS3 is the best which is followed by PS and FS1.

\begin{table}[!htbp]
\caption{The average of RPEs for Eye data set.}
\label{Tab:eyedata}
\centering
{%
\begin{tabular}{lcccccccc}
\toprule
ALASSO
& FS1
& FS3
& PS \\
\midrule
1.0598 &1.0423 &1.0626& 1.0430  \\
\bottomrule
\end{tabular}%
}
\end{table}

\subsection{Riboavin Data}
Here, we consider the  data set about riboavin (vitamin B2) production
in Bacillus subtilis. There is a single real valued
response variable which is the logarithm of the riboavin production rate. Furthermore,
there are $p=4088$ explanatory variables measuring the logarithm of the expression level of
4088 genes. There is one rather homogeneous data set from $n = 71$
samples that were hybridized repeatedly during a fed batch fermentation process where
different engineered strains and strains grown under different fermentation conditions
were analyzed.

\begin{figure}[!htbp]
\centering
\includegraphics[width=15cm,height=8cm]{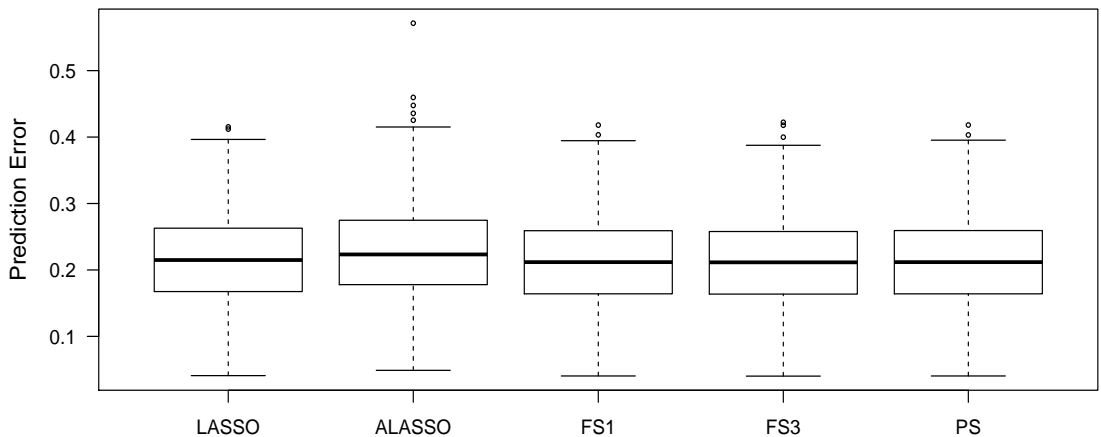}
\caption{The average of RPEs for Riboflavin data set.
 \label{Tab:riboavin}}
\end{figure}

For this data, LASSO and ALASSO select 27 and 12 significant covariates, respectively. Notice that we used ``one standard error'' rule for selection of tuning parameters. We did not scale the design matrix and include the intercept term. The result of RPEs is shown in Table \ref{Tab:riboavin} and Figure \ref{Fig:riboavin:box}.  Again, it is clear that the performance of FS3 outshines PS and FS1 even though the performance of ALASSO is less efficient than LASSO.

\begin{table}[!htbp]
\caption{Relative Prediction Error of estimators}
\label{Fig:riboavin:box}
\centering
{%
\begin{tabular}{lccccc}
\toprule

ALASSO
& FS1
& FS3
& PS \\
\midrule
	
0.9509 &1.0222& 1.0273& 1.0225  \\
\bottomrule
\end{tabular}%
}
\end{table}

\section{Conclusions}
In this paper, we extended variable selection methods to a new direction to be shrunken to a targeted estimator. Specifically we combined estimation strategies from both under-fitted and over-fitted models, in a high-dimensional regression model, employing a bounded measurable function. Specific concave functions were adopted to show the superiority of the proposed double shrunken estimators over the best of \citet{AhmedYuzbasi2016}. We have conducted a simulation study to investigate the performance of the suggested shrinkage strategy with respect to two penalty estimators: LASSO and ALASSO. According to the simulation results, the performance of under-fitted estimator ALASSO is the best since it is based on true model, and the FS3 performs better than both FS1 and PS when $\Delta=0$. On the other hand, the RMSE of the ALASSO decreases and approaches to zero while the all others approach to one when we increase the magnitude of weak signals. We further analysed two high-dimensional data sets, and the performance of the shrinkage strategy was striking.

We also proposed another shrinkage estimator, which was not included in the numerical analyses, namely FS2 given by
\begin{eqnarray*}
\boldsymbol{\widehat{\beta}}_1^{\rm FS2}&=&\boldsymbol{\widehat{\beta}}_1^{\rm UF}+\left( \boldsymbol{\widehat{\beta}}_1^{\rm OF}-\boldsymbol{%
\widehat{\beta}}_1^{\rm UF}\right) \left( 1-\frac{(p_{2}-2)\exp(-W_n^2)}{W_{n}}\right) \text{, }p_{2}\geq 3,
\end{eqnarray*}
We evaluated the performance of FS2 comparatively. We realized, it is not competitor to FS1, FS3 and PS, specially for small values $\Delta$. However, as soon as the non-centrality parameter $\Delta$ gets larger, its RMSE goes to 1 dominating all other estimators uniformly, since the weight goes to infinity and it simplifies to over-fitted model.


\end{document}